\documentstyle [12pt,epsf]{article}

\input{epsf}
\begin{document}
\begin{titlepage}
\begin{center}

\vspace{-0.4in}

{\Large \bf The Composite Operator (CJT) Formalism in the \\
$(\lambda\varphi^4+\eta\varphi^{6})_{D=3}$ Model at Finite Temperature}\\
\vspace{.6in}
{\large\em G.N.J.A\~na\~nos\footnotemark[1], N.F.Svaiter\footnotemark[2]}\\
Centro Brasileiro de Pesquisas F\'{\i}sicas-CBPF\\ Rua Dr.Xavier
 Sigaud 150, Rio de Janeiro, RJ 22290-180 Brazil\\

\subsection*{Abstract}
\end{center}

We discuss three-dimensional $ \lambda\varphi^4+\eta\varphi^6 $ theory in the 
context of the $ 1/N $ expansion at finite temperature. 
We use the method of the composite operator (CJT) for summing a large sets
 of Feynman graphs. We analyse the behavior of the thermal  
square mass and the  thermal coupling constant in the low and high temperature
limit. The existent of the tricritical point at some temperature is found 
using this non-pertubative method.

\footnotetext[1]{e-mail:gino@lafex.cbpf.br}
\footnotetext[2]{e-mail:nfuxsvai@lafex.cbpf.br}

\end{titlepage}

\newpage

\baselineskip .37in
\section {Introduction}

The conventional perturbation theory in the coupling or loop expansion can only
be used for the study of small quantum corrections to classical results.
When discussing truly quantum mechanical effects to any given order in such
an expansion, one is not usually able to justify the neglect of yet higher
 order. In other way, for theories with a large $N$ dimensional internal 
symmetry group there exist another perturbation scheme, the $1/N$ expansion, which circumvents this criticism. Each term in the $ 1/N $ expansion 
contains an infinite subset of terms of the loop expansion.
The $ 1/N $ expansion has the nice property that the leading-order quantum
 corrections are the same order as the classical quantities. Consequently, the
leading order which adequately characterizes the theory in the large $N$ limit
preserves much of the nonlinear structure of the full theory.
In the next section we derive the effective action to leading order in $1/N$
for three dimensions and consequently the effective potential. 
In this paper we compute effective thermal mass which depend on the 
temperature for the field theory with $\varphi^6$ interaction on $D=3$ 
Euclidean dimensions.
Its is known that, in $D>4$, such theories with $\varphi^4$ interaction are
in fact free field theory, while in $D<4$ they have a non-trivial continuum
limit  as an interacting field theory. 

Since both the six-point coupling of $(\eta\varphi^6)_{D=3}$ and the four-point
coupling of $(\lambda \varphi^4)_{D=4}$ are dimensionless one expected that have the same continuum limit.
However it has been shown that, in the large $N$ limit, the
 $(\eta\varphi^6)$ theory has a UV fixed point for $D=3$ and therefore must have a second infra-red fixed point \cite{bardeen}. At least for large $N$  the $\varphi^6$ theory is known to be qualitatively different from $(\lambda \varphi^4)_{D=4}$ theory. We study the large N expansion, using the methods of composite operator \cite{cornwall}. The organization of the letter is the following. In section
II we briefly discuss the composite operator (CJT) formalism. In section
III the thermal gap equation is derived. In section IV the tricritical
phenomenon is presented. Conclusions are given in section V. In this 
paper we use $c=k_{B}=\hbar=1$.

\section{The effective Potential( The CJT Formalism)}

We are interested here in the most general renormalizable 
scalar field theories $\lambda\varphi^{4}+\eta\varphi^{6}$ 
possessing an internal symmetry $ O(N)$, in three 
dimensions. For simplicity we will call a $\varphi^{6}$ model. 
We use the method of composite operator developed by Cornwall, Jackiw and Tomboulis \cite{cornwall,livro} for summing large sets of Feynman graphs by considering only two-particle irreducible (2PI) graphs express in terms of the exact propagators. This technique lead to the formulation of the
effective action and 
effective potential which is functional the 
vacuum expectation value of both the quantum field $ \varphi(x) $ 
and the time ordered product $ T\{\varphi(x)\varphi(y)\} $ of the fields. 
Using this method 
Townsend derived the effective potential of $\varphi^6$ theory to leading order in the $\frac{1}{N}$ expansion for $D \leq 3$ \cite{townsend} 
and  proof that $\frac{1}{N}$ expansion is consistent for $\varphi^6$ to leading order in $\frac{1}{N}$.  
The Lagrangian density of the $O(N)$ symmetry 
$\varphi^6$ theory is :
\begin{equation}
{\cal L}(\varphi)=\frac{1}{2} (\partial_{\mu} \varphi)^2-\frac{1}{2}m_{0}^2 
\varphi^2-\frac{\lambda_0}{4 N!} \varphi^4- \frac{\eta_0}{6! N^2}\varphi^6, 
\end{equation}   
where the quantum field is an $N$-component vector 
in the $N$-dimensional
internal symmetry space.
For definiteness, we work at zero-temperature; however, 
the finite temperature
generalizations can be easily obtained \cite{dolan}. 
We are interested in the effective action $\Gamma (\phi)$ which governs 
the behavior of the expectation values $\varphi_a (x)$ of the quantum field 
where $\phi$ is given by
\begin{equation}
\phi(x) \equiv {\delta W(J) \over \delta J(x)} = <0|\varphi|0>,
\end{equation}
where $W(J)$ is the generating functional for connected Green's functions.

$\Gamma (\phi)$ can be show to be the sum of one-particle irreducible (1PI)
Feynman graphs with a factor $\phi_a (x)$ on the external line.
We may use of the formalism of composite operator who reduce  the problem 
 to summing two particle irreducible (2PI) Feynman graphs by defining a 
generalized effective action $\Gamma (\phi,G)$ which is a functional not only 
 $\phi_a (x)$, but also of the expectation values $G_{ab} (x,y)$ of the
time ordered product of quantum fields $ T\{\varphi(x)\varphi(y)\} $.
\begin{equation}
 \Gamma (\phi,G)= I(\phi)+\frac{i}{2} Tr \ln G^{-1} + 
\frac{i}{2}   Tr D^{-1}(\phi) G + \Gamma_2(\phi,G) +\dots \;\;.
\label{G1}
\end{equation}
where $I(\phi)=\int dx^D {\cal L}(\phi) $ , $G$ is a shorthand for the $G_{ab}(x,y)$ and  $D$ is defined by 
\begin{equation}
i D^{-1}=\frac{\delta^2 I(\phi)}{\delta\phi(x) \delta\phi(y)},
 \end{equation}         
and is shorthand for $D_{ab}(\phi;x,y)$, $\Gamma_2(\phi,G)$ is computed as
follows. In the classical action $I(\varphi)$ shift the field $\varphi$ by 
$\phi$. The new action  $I(\varphi+\phi) $  posses term cubic and higher in 
$\varphi$; this define an interaction part   $I_{int} (\varphi,\phi)$
where the vertices depend on $\phi$.  $\Gamma_2(\phi,G)$ is given by 
sum of all  (2PI) vacuum graphs  in a theory with vertices determined by $I_{int} (\varphi,\phi)$ and the propagators set equal to 
$G(x,y)$. The trace and logarithm in eq.(\ref{G1}) are functional.
\begin{eqnarray}
 {\cal L}_{int}(\varphi,\phi)&=&-\frac{1}{2} \left( \frac{\lambda_0 \phi_a}{3N} + \frac{\eta_0 \phi^2\phi_a}{30N^2} \right) \varphi_a \varphi^2 -
 \left( \frac{8\eta_0 \phi_a\phi_b\phi_c}{6N^2} \right) \varphi_a \varphi_b
\varphi_c -\frac{1}{4!N} \left( \lambda_0+ \frac{\eta_0\phi^2}{10N}\right)
 \varphi^4 \nonumber \\
& & 
-\left( \frac{12\eta_0\phi_a\phi_b}{6!N^2} \right)\varphi_a\varphi_b\varphi^2
-\frac{1}{5!}\left( \frac{\eta_0\phi_a}{N^2}\varphi_a\varphi^4\right)
-\frac{\eta_0}{6!N^2}\varphi^6.
\end{eqnarray}
The effective action as usually defined is found by solving for $G_{ab}(x,y)$ in the equation 
\begin{equation}
\frac{\delta \Gamma(\phi,G)}{\delta G_{ab}(x,y)}=0
\label{eq1}
\end{equation}
and substituting in the generalized effective action $\Gamma(\phi,G)$.

The vertices in the above equation contains factor $1/N$ or $1/N^2$, but 
a $\varphi$ loop gives a factor of N provided the $O(N)$ isospin flows
around it alone and not into another part of the graph. We usually call
such loops bubbles. Then the leading order in $1/N$ the vacuum graphs are
bubbles trees with two or three bubbles at each vertex. The (2PI) graphs
are shown in figure.(1).
\begin{figure}[ht]
  
\centerline{\epsfysize=1.0in\epsffile{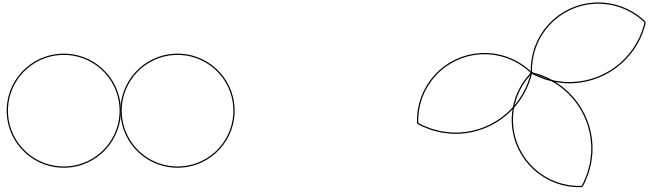}} 
\caption[region]
{The 2PI vacuum graphs }
\end{figure}
It is straightforward to obtain
\begin{equation}
\Gamma_2(\phi,G)=\frac{-1}{4!N}\int d^{D}x \left( \lambda_0+
\frac{\eta_0\phi^2}{10N} \right) [G_{aa}(x,x)]^2 -
\frac{\eta_0}{6!N^2}
\int d^{D}x [G_{aa}(x,x)]^3 .
\label{eq2}
\end{equation}

Therefore eq.(\ref{eq1}) becomes

\begin{eqnarray}
\frac{\delta\Gamma(\phi,G)}{\delta G_{ab}(x,y)}&=& \frac{1}{2}(G^{-1})_{ab}
(x,y) + \frac{i}{2} D^{-1}(\phi)-\frac{1}{12N}\left( \lambda_0+
\frac{\eta_0\phi^2}{10N}\right)[\delta_{ab}G_{cc}(x,x)]\delta^D(x-y)
\nonumber \\
& & -\frac{3\eta_0}{6!N} \delta_{ab}[G_{cc}(x,x)]^2\delta^D(x-y)=0.
\label{eq3}
\end{eqnarray} 
Rewriting this equation , we obtain the gap equation
 \begin{eqnarray}
(G^{-1})_{ab}(x,y) &=& D^{-1}(\phi)+ \frac{i }{6N}\left( \lambda_0+
\frac{\eta_0\phi^2}{10N}\right)[\delta_{ab}G_{cc}(x,x)]\delta^D(x-y)+\nonumber
\\
 & &\frac{i\eta_0}{5!N^2} \delta_{ab}[G_{cc}(x,x)]^2\delta^D(x-y).
\label{eq4}
\end{eqnarray}
Hence 
\begin{equation}
\frac{i}{2} Tr D^{-1}G= \frac{1}{12N}\int d^Dx \left( \lambda_0+\frac{\eta_0\phi^2}{10N} \right) [G_{aa}(x,x)]^2 + 
\frac{2\eta_0}{6!N^2}\int d^{D}x [G_{aa}(x,x)]^3+ cte .
\label{eq5}
\end{equation}
Using eqs. (\ref{eq4}) and (\ref{eq5}) in (\ref{eq2}) we find the effective action
\begin{eqnarray}
\Gamma(\phi) &=& I(\phi)+\frac{i}{2} Tr \ln G^{-1}+\frac{1}{4!N}
\int d^Dx \left( \lambda_0+\frac{\eta_0\phi^2}{10N} \right) [G_{aa}(x,x)]^2+
 \nonumber \\
& & \frac{2\eta_0}{6!N^2} \int d^Dx [G_{aa}(x,x)]^3 ,
\label{effa}
\end{eqnarray}
where $G_{ab}$ is given implicitly by eq.(\ref{eq4}).
The last two term on the r.h.s of eq.(\ref{effa}) is the leading two-loop 
contribution to the effective action for composite operators in  $\lambda \phi^4+ \eta\phi^6$ theory.
As usual we may simplify by separating $G_{ab}$ into transverse and
longitudinal components, so 
\begin{equation}
G_{ab}=(\delta_{ab}-\frac{\phi_a \phi_b}{\phi^2})g+\frac{\phi_a \phi_b}{\phi^2} \stackrel{\sim}{g}\; ,
\end{equation}
in this form we can invert $G_{ab}$,
\begin{equation}
(G)^{-1}_{ab}=(\delta_{ab}-\frac{\phi_a \phi_b }{\phi^2} ) g^{-1} +
\frac{\phi_a \phi_b}{\phi^2} {\stackrel{\sim}{g}}^{-1}\; .
\end{equation}

Now we can take the trace respect the indices,
\begin{equation}
G_{aa}=Ng +O(1) ,\;\;\;\; (G)^{-1}_{aa}=Ng^{-1}+ O(1) \; .
\label{gd}
\end{equation}
From this equation to leading order in $\frac{1}{N}$, $G_{ab}$ is diagonal in $a,b$. Substituting eq.(\ref{gd}) into eq.(\ref{effa}) and
eq.(\ref{eq4}) and keep only the leading order  one finds  the daisy and superdaisy resumed effective potential for the $\lambda \phi^4+ \eta\phi^6$ theory is given by:
\begin{eqnarray}
\Gamma(\phi)&=&I(\phi)+\frac{iN}{2} Tr \ln g^{-1}+\frac{N}{4}
\int d^Dx \left( \lambda_0+\frac{\eta_0\phi^2}{10N} \right) g^2(x,x)+
\nonumber \\
 & & \frac{2N\eta_0}{6!} \int d^Dx g^3(x,x) +O(1) ,
\label{effa1}
\end{eqnarray}
and the gap equation
\begin{equation}
g^{-1}(x,y)=i\left[ \Box + m_0^2+
\frac{\lambda_0}{6}(\frac{\phi^2}{N}+ g(x,x)) + \frac{\eta_0}{5!}
(\frac{\phi^2}{N}+ g(x,x))^2 \right ] \delta^{D}(x-y)+ O(\frac{1}{N})
\label{gap1}
\end{equation}
Now it is convenient to continue all momenta to Euclidean values $(p_0=ip_4)$
and take the following Ansatz for $g(x,y)$,
\begin{equation}
g(x,y)=\int \frac{d^{D}p}{(2\pi)^D}\frac{\exp^{i(x-y)p}}{p^2+M^2(\phi)}
\end{equation}
and substituting this in eq.(\ref{gap1}) we get the expression for 
the gap equation:  
\begin{equation}
 M^2(\phi)=m^2_0+
\frac{\lambda_0}{6}\left( \frac{\phi^2}{N}+F(\phi)\right) + \frac{\eta_0}{5!} \left( \frac{\phi^2}{N}+ F(\phi)\right)^2,
\label{M2}
\end{equation}
where $F(\phi)$ is given by
\begin{equation}
F(\phi)=\int \frac{d^Dp}{(2\pi)^D}\frac{1}{p^2+M^2(\phi)},
\label{Fphi}
\end{equation}
and the effective potential in Euclidean-space can be express, 
\begin{equation}
Veff(\phi)=V_0(\phi)+\frac{N}{2}\int \frac{d^Dp}{(2\pi)^D} \ln \left [p^2+M^2(\phi) \right ] 
 -\frac{N}{4!}(\lambda_0+\frac{\eta_0 \phi^2}{10N}) F(\phi)^2-
\frac{2N\eta F(\phi)^3}{6!} .
\label{ve1}
\end{equation}
where $V_0(\phi)$ is the classical potential.

\section{The Gap-Equation for the model $(\lambda \varphi^4+ \eta\varphi^6)_{D=3}$ at Finite Temperature}

Let us suppose that our system is in equilibrium with a thermal bath.
At the one-loop approximation the thermal mass and coupling 
constant for the $\lambda\varphi^{4}$ model in a D-dimensional 
Euclidean space have been obtained in a previous work \cite{adolfo} and 
for the the theory $(\lambda \varphi^4)_{D} $ is obtained using 
the composite operator method 
\cite{ananos}. It is clear that for $D=3$ the $(\lambda \phi^4)_{D}$ model
is not the most general perturbative renormalizable model 
and we are able to introduce  a $\phi^6$ term
preserving the renormalizability of theory. For $D=3$  the  $\lambda \varphi^4+ \eta\varphi^6$ model is the most general O(N) symmetric model
which preserves the perturbative renormalizability. Most general 
models with cubic symmetry can be also studied. In the conclusion
we will briefly discuss this model.    

 To study the temperature effects in quantum field theory the most used methods is the imaginary time Green function approach \cite{dolan}, which amounts to replace the 
continuous four momenta $k_4$ by discrete $\omega_n$ an integration by a 
summation ($\beta=\frac{1}{T}$):
\begin{eqnarray}
 k_4 &\rightarrow& \omega_n =\frac{2\pi n}{\beta}, \;\;\;\;\; n=0,\pm1,\pm2,... 
\nonumber \\
\int\frac{d^Dk}{(2\pi)^D} &\rightarrow &  \sum_{n} \frac{1}{\beta}\;\; \int\frac{d^{D-1}k}{(2\pi)^{D-1}} .
\end{eqnarray}

Then the gap equation at  finite temperature  for the theory $\lambda \varphi^4+ \eta\varphi^6$ is  given by (see eq.(\ref{M2})),
\begin{equation}
 M^2_{\beta}(\phi)=m^2_0+
\frac{\lambda_0}{6}\left( \frac{\phi^2}{N}+F_{\beta}(\phi)\right) + \frac{\eta_0}{5!} \left( \frac{\phi^2}{N}+ F_{\beta}(\phi)\right)^2,
\label{M2T}
\end{equation}
where
\begin{equation}
F_{\beta}(\phi)=\frac{1}{\beta}\sum_{n=-\infty}^{\infty} \int \frac{d^{D-1}p}{(2\pi)^{D-1}} \frac{1}{\omega^2_n+p^2+M^2_{\beta}(\phi)}\; .
\label{Fbphi}
\end{equation}
In order to regularized this expression we use a mixing between dimensional regularization and analytic regularization. For this purpose we define the 
expression $I_{\beta}(d,s)$ as :
 \begin{equation}
I_{\beta}(D,s,m)=\frac{1}{\beta}\sum_{n=-\infty}^{\infty}\int
\frac{d^{D-1}k}{(2\pi)^{D-1}}\frac{1}{(\omega^2_n+k^2+m^2)^s} \; .
\end{equation}
Using analytic extension of the inhomogeneous Epstein zeta function
it is possible to obtain the analytic extension of $I_{\beta}(D,s,m_{\beta})$;
\begin{equation}
I_{\beta}(D,s,m)= \frac{m^{D-2s}}{(2\pi^{1/2})^D\Gamma(s)} \left[
\Gamma(s-\frac{D}{2}) + 4 \sum_{n=1}^{\infty} \left( \frac{2}{mn\beta}
 \right)^{D/2-s} K_{D/2-s}(mn\beta) \right]
\end{equation}
where $K_{\mu}(z)$ is the modified  Bessel function of third kind.
Fortunately for $D=3$ the analytic extension of the function $I_{\beta}(D,s,m_{\beta})$ is finite and can be express in
a closed form \cite{an} (note in $D=3$ we have no pole, at least in this 
approximation), and particular 

\begin{equation}
F_{\beta}(\phi)=I_{\beta}(3,1,M_{\beta}(\phi))=-\frac{M_{\beta}(\phi)}{4\pi}\left( 1+\frac{2\ln(1-e^{-M_{\beta}(\phi)\beta})}{M_{\beta}(\phi)\beta}
\right ).
\label{m2b2}
\end{equation}
If we have no spontaneous symmetric breaking, we define the thermal  effective
mass as 
\begin{equation}
\left. \frac{\partial^2 Veff(\phi)}{\partial \phi_a^2}\right|_{\phi=0}= 
m^2_{\beta} 
\end{equation}
and is not difficult to show that,
\begin{equation}
m^2_{\beta}=M^2_{\beta}(0)=m^2+
\frac{\lambda}{6}F_{\beta}(0)+\frac{\eta}{5!}F_{\beta}(0)^2
\label{m2beta}
\end{equation}
and  
\begin{equation}
F_{\beta}(0)=-\frac{m_{\beta}}{4\pi}\left( 1+\frac{2\ln(1-e^{-m_{\beta}\beta})}{m_{\beta}\beta}
\right ).
\label{Fbeta}
\end{equation}
where the $m^2$,$\lambda$ and $\eta$ are the finite parameter of the 
theory at tree-level. 
So eq.(\ref{m2beta}) and eq.(\ref{Fbeta}) determined the behavior of
the  effective mass with the temperature and from this relations it is
possible to show that the mass increase with the temperature.
 
And the expression for the effective coupling constant can be derived from
the relation given  by:
\begin{equation}
\left. \frac{\partial^4 Veff(\phi)}{\partial \phi_a^4}\right|_{\phi=0}=\left. 3\frac{\partial^2 M^2_{\beta}(\phi)}{\partial^2 \phi_a}\right|_{\phi=0} 
\end{equation}
where $M^2_{\beta}(\phi)$ is given by eq.(\ref{M2}), so after some algebra 
we find that,
\begin{equation}
\lambda_{\beta}=\frac{\lambda +\frac{\eta}{10}  
 F_{\beta}(0)} 
{1-\left[ \frac{\lambda_0}{6}+\frac{2\eta_0 }{5!} F_{\beta}(0) \right]
 \frac{\partial F_{\beta}(0)}{\partial m^2_{\beta}} };
\label{lb}
\end{equation}
This is the thermal effective coupling constant at leading order.
In order to found the behavior of the thermal effective  coupling constant with the temperature, we used the solution for the eq.(\ref{m2beta}) and replace  in eq.(\ref{lb}). From this we conclude that the effective thermal coupling constant increase with the temperature.
In the next section we discuss the existence of the 
tricritical phenomenon in this  model at finite temperature.

\section{The tricritical phenomenon}
In the last section we obtained the thermal correction to the square mass 
$m^2_{\beta}$ an the coupling constant $\lambda_{\beta}$ in absence of 
spontaneous symmetry breaking.  The tricritical phenomenon occurs when  $\lambda_{\beta}=m^2_{\beta}=0$. If it happens we conclude that spontaneous symmetry breaking must be  occur. At tree level this happen when the classical potential $V_0(\phi)$, develops an absolute minimum for $\phi^2 \neq 0$. If we consider 
quantum effects, then from eq.(\ref{ve1}) we have the relation(we discus at
$T=0$) 
\begin{equation}
\frac{\partial V}{\partial \phi_a}=\left[ m^2+\frac{\lambda}{6}
\left( \frac{\phi^2}{N}+F(\phi) \right) + \frac{\eta}{5!}\left( \frac{\phi^2}{N}+F(\phi) \right)^2 \right] \phi_a 
\end{equation}
If we have spontaneous symmetry breaking the next relation must be satisfied in terms of the renormalized parameter,
\begin{equation}
\left[ m^2+\frac{\lambda}{6}
\left( \frac{\phi^2}{N}+F(\phi) \right) + \frac{\eta}{5!}\left( \frac{\phi^2}{N}+F(\phi) \right)^2 \right]=0 \; .
\label{ssb}
\end{equation}
But form eq.(\ref{Fphi}) $F(\phi)$ can be express as (using dimensional regularization)
\begin{equation}
F(\phi)=-\frac{1}{4\pi}\left[ m^2+\frac{\lambda}{6}
\left( \frac{\phi^2}{N}+F(\phi) \right) + \frac{\eta}{5!}\left( \frac{\phi^2}{N}+F(\phi) \right)^2 \right]^{\frac{1}{2}}.
\end{equation}
From the last two relations we conclude that $F(\phi)=0$ at the minimum.
This implies that eq.(\ref{ssb}) becomes 
\begin{equation}
 m^2+\frac{\lambda}{6}
\left( \frac{\phi^2}{N}\right) + \frac{\eta}{5!}\left( \frac{\phi^2}{N}\right)^2 =0 \; .
\label{ssb1}
\end{equation}
This is the same equations as for the tree approximation, but in terms of
the renormalized parameters. 
And the effective  square mass vanishes at the minimum i.e $M^2(\phi)=0$, this
is consequence of the Goldstone`s theorem.  
From eq.(\ref{ssb1}) provides tree possibilities for symmetry breaking :
\begin{equation}
\begin{array}[pos]{lll}
 1. \; & \lambda \geq 0 \; , & m^2 < 0\\
 2. \; & \lambda < 0 \; , & m^2 < 0\\ 
 3. \; & \lambda < 0 \; , & m^2 \geq 0 .
 \end{array}
\end{equation}
\begin{figure}[ht]
  
\centerline{\epsfysize=3.0in\epsffile{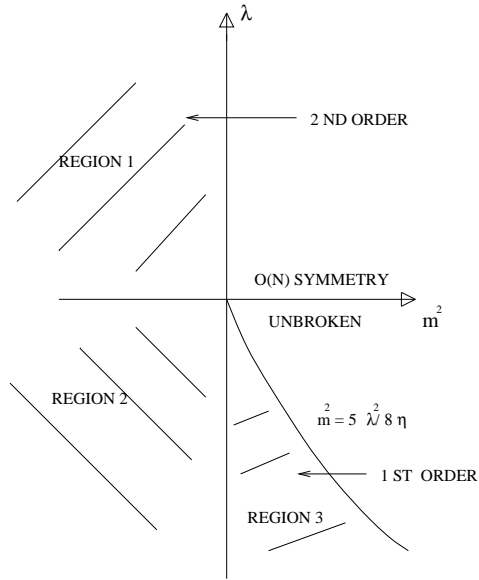}} 
\caption[region]
{The Regions in $\lambda$ , $m^2$ of symmetry breaking for fixed $\eta$}
\end{figure}
If we consider the region 3 at the tree level and for $ |\lambda|$ 
\begin{equation}
\frac{5\lambda^2}{8\eta} \; > \; m^2 \; .
\label{cons1}
\end{equation}
sufficiently large we have  two real positive zeros for $V_0(\phi)$.
In this case for $m^2>0$ at $\frac{5\lambda^2}{8\eta}=m^2$ we have a first order transition as shown in figure.(2). Then the point  $m^2=0$,  $\lambda=0$ is the tricritical point, which  the first and second order  transitions meet here. Of course the relation in eq.(\ref{cons1}) was found by requiring that 
the minimum of the classical potential in region 3 be absolute. If we consider 
quantum corrections this relation must be modified.
At finite temperature the effective parameter $m^2$ and $\lambda$ change with the temperature as we discussed in the preceding section. At sufficiently high temperature we guess form eq.(\ref{lb}) that there is no  symmetry breaking. Then in order to use the relations of the last  section we first started when the symmetry is present and from this we prove that it can be exist an intermediate temperature $\beta_c$ at which  $ m_{\beta_c}^2 =0 $,  $\lambda_{\beta_c}=0$ .

First we note that the eq.(\ref{m2b2}) is valid only for $m^2_{\beta} > 0 $ and this expression is not useful to find the critical temperature. So we 
regularized this expression in the limit $m^2_{\beta}=0,\; (m^2_{\beta}>0)$ and for $D=3-\epsilon$ we get,
\begin{equation}
\begin{array}{ll}
\lim\;\;I_{\beta}(3-\epsilon,s=1,m_{\beta}) & =\frac{\Gamma(\frac{1-\epsilon}{2})\; \zeta(1-\epsilon)}{2 (\pi)^{3/2-\epsilon/2}\beta_c^{1-\epsilon}} \\
m_{\beta}\rightarrow 0 \;\; (m_{\beta}>0) &  
\end{array}
\end{equation}
the finite part $Ir({\beta_c})$ given by (we use the expansion of $\zeta(z)$ an $\Gamma(z) $ function in order to find the regular part), 
\begin{equation}
Ir(\beta_c)=\frac{1}{2\pi\beta_c}\ln(\frac{k}{\pi\beta_c^2\mu_c^2});
\end{equation}
where $\mu_c \approx m$ is the mass parameter and $k$ is a numeric constant. The
infrared-induced uncertainties can be subsumed into single parameter  
$\mu_c$ \cite{einhorn}.  
So the tricritical phenomenon occurs when
\begin{eqnarray}
m^2_{\beta} &=& 0=m^2+
\frac{\lambda}{6}Ir(\beta_c) + \frac{\eta}{5!}Ir^2(\beta_c) ,
\nonumber \\
\lambda_{\beta} &=&0=\lambda+\frac{\eta}{10}Ir(\beta_c).
\end{eqnarray} 
And from this relations we conclude that if the tricritical phenomena occurs
we must be keep the next two following relation:
\begin{equation}
\begin{array}{ll}
\lambda^2= & \frac{6\eta m^2}{5} \\
T_c^2= & \frac{\mu_c^2\pi}{k}\exp{(\frac{-\lambda\pi}{5\eta T_c})}.
\end{array}
\label{Tc}
\end{equation}
Where $T_c$ is the critical temperature at which the tricritical phenomenon occurs. Of course the parameters $(m^2,\lambda)$ must be in the region 3
in order to occur the tricritical phenomena an obey the relation eq.(\ref{Tc}).
\section{Conclusions}
We have done in this paper an analysis of the vector model
$\lambda\varphi^{4}+ \eta\phi^6$  in $D=3$ Euclidean 
dimensions at finite temperature. 
The form of the thermal corrections to the mass 
and coupling constant have been found using resummation methods in the 
leading order $\frac{1}{N}$ approximation (Hartree-Fock approximation).
We conclude that these parameters increase with the temperature. This ressult is consistent with previous work  \cite{ananos} in the sense
that we are able to recover in the limit $\eta=0$. 

We discussed the existent of the tricritical phenomenon at finite temperature  and we have been found an  expression for the critical temperature 
at which the thermal effective mass  and coupling constant vanishes, and this shown that the tricritical phenomena occurs at an intermediate temperature $T_c$.

A natural extension of this work is to study the cubic anisotropic model with the following Lagrange density
\begin{equation}
{\cal L}_{int}=
\lambda_1\sum_{i=1}^{N} (\phi_i\phi_i)^2+\eta_1\sum_{i=1}^{N} (\phi_i\phi_i)^3
+\lambda_2\sum_{i=1}^{N} (\phi_i)^4 +\eta_2\sum_{i=1}^{N} (\phi_i)^6.
\end{equation}
In $D=4$ with the $\lambda\phi^{4}$ model appear different fixed 
points (a Gaussian, Heisenberg, Ising and cubic) \cite{Amit}
and between them the Heisenberg fixed point is the 
only stable when $M<4$. For $M>4$ the cubic fixed point becomes 
stable. This model is under investigation by the authors.
    
\section{Acknowlegements}

This paper was supported by Conselho Nacional de 
Desenvolvimento Cientifico e Tecnologico do Brazil (CNPq).

\end{document}